\documentclass[10pt,a4paper,aps,twocolumn,showpacs,superscriptaddress,nofootinbib]{revtex4-1}
\usepackage[colorlinks,hyperindex]{hyperref}
\usepackage{epsfig}
\usepackage{float}
\usepackage{float,upgreek,yfonts}
\RequirePackage{amssymb,amsmath}
\usepackage[latin1]{inputenc}
\usepackage{graphicx}
\usepackage{indentfirst}
\usepackage{color}
\begin{document}
\title{Hadron multiplicity calculation: a configurational entropy approach to the saturation scale in QCD}
\author{G. Karapetyan}
\email{gayane.karapetyan@ufabc.edu.br}
\affiliation{Centro de Ci\^encias Naturais e Humanas, Universidade Federal do ABC - UFABC, 09210-580, Santo Andr\'e, Brazil}
\begin{abstract}
This paper investigates the configurational entropic content of hadron-nucleus collisions. Hadron multiplicities and $Au$ nuclei are employed
to compute the critical points of the configurational entropy as a function of the saturation scale in deep inelastic scatterings, in QCD. The results match phenomenological data to the precision of 0.39\%.

\end{abstract}
\maketitle
\section{Introduction}

The concept of the configurational entropy (CE) was introduced properly to find the correct value of the experimental parameters and data in the framework of QCD and Color Glass Condensate approximation \cite{Karapetyan:2018yhm}. Such a technique nowadays can straightforwardly be utilized as a trustworthy apparatus in the study of different channels of hot nuclear interaction \cite{Karapetyan:plb}.
This approach has been most frequently used to study the stability of  nuclear systems at high excitations, as well as the dominance of quantum resonances that correspond to critical points of the CE. One can also use the critical points of the CE, underlying the evolution of  interactions.
The critical points of the CE   have already been  extensively investigated in QCD and particle physics. In fact, scalar and tensor mesons \cite{Ferreira:2019inu,Bernardini:2018uuy,Braga:2018fyc,Bernardini:2016hvx,Barbosa-Cendejas:2018mng,Ferreira:2019nkz}, glueballs \cite{Bernardini:2016qit}, charmonium and bottomonium  \cite{Braga:2017fsb}, the quark-gluon plasma \cite{daSilva:2017jay}, barions \cite{Colangelo:2018mrt} and several other systems in QCD have been scrutinized \cite{Ma:2018wtw}, using AdS/QCD and information entropy.
The CE, founded in Refs. \cite{Gleiser:2012tu,Gleiser:2011di, Gleiser:2013mga}, emulating the Shannon's information entropy,
was also studied in Refs.
\cite{Gleiser:2018kbq,Sowinski:2015cfa,Gleiser:2014ipa,Gleiser:2015rwa}.
In addition, it was used in order to predict the relative stability of physical configurations in Refs. \cite{Casadio:2016aum,Fernandes-Silva:2019fez,Braga:2016wzx}, for AdS black holes and their quantum portrait as Bose--Einstein graviton condensates. Skyrmions, emulating magnetic structures, were scrutinized in Ref. \cite{Bazeia:2018uyg}. Other physical aspects of the CE were explored in Refs. \cite{Bernardini:2019stn,Alves:2017ljt,Alves:2014ksa}.

The CE approach to the Color Glass Condensate
(CGC) has spectacular results in the high energy regime, matching, corroborating and predicting phenomenological and experimental data  \cite{Karapetyan:2018yhm,Karapetyan:2019epl,Ma:2018wtw,Karapetyan:plb,Karapetyan:2017edu,Karapetyan:2016fai}.
At such regime, the parton saturation can be observed. This is caused by strong coherent
gluon fields \cite{Kharzeev}.
The LHC open the possibility to explore QCD, in order to deeply investigate and study several new experimental results  in the high density parton mode. General features of the inelastic nucleus-nucleus or/and proton-proton collisions at LHC energy range
can be well described in the frame of CGC.
To describe the data from RHIC, the calculation at the low $x$ regime can be considered.
One of the main features of the computations involves the multiplicities of partons.
As it has been suggested in Ref. \cite{Kharzeev}, the above mentioned features do not vary substantially from the initial channel of the interaction up to its final state. It can thus be interpreted as a local parton hadron duality, in other words, the entropy conservation.
Obtaining these features of the interaction
can enlight more deeply the mechanism and the dynamics of the collision \cite{Karapetyan:2018yhm,Karapetyan:plb}.
It is interesting to use, instead of the energy density of any spatially localized system such a quantity as the reaction cross section, which is also spatially localized, defining the nuclear CE. In such a case, the Fourier transform of the reaction probability allows to derive the critical points of the nuclear CE \cite{Karapetyan:2018yhm,Karapetyan:plb,Karapetyan:2017edu,Karapetyan:2016fai}.
Another possibility, to be employed here, is to compute the CE using the hadronic multiplicity as the localized function, for fixed values of the rapidity.

In this context, the critical points of the CE, underlying some system in QCD, can benchmark the existing experimental data and also to find the most convenient parameters that can describe several nuclear phenomena.
Using the CE concept we provide the saturation scale parameter, which predicts the value of the hadron multiplicity at LHC, based on the CGC approach.

The present paper is presented as follows.
In the second section, we present the general formalism of
nucleus-nucleus or/and hadron-nucleus collisions.
The third section is devoted to giving some details of the hadron multiplicities and on the influence of higher-order corrections and the effects of the coupling constants on the obtained results. We present the results and how the critical point of the CE determine an important physical parameter that defines  the saturation scale in the case of deep inelastic scatterings. We then summarize our results in the conclusion.

\section{Nucleus--nucleus collisions in the frame of Glauber approach}

During collisions of high-energy nucleons, the track nucleons are  assumed to go into a straight line, due to a small scattering angle as well as the small radius for the nucleon-nucleon collision.
Nucleons in interaction can be classified as participants,  $N_{\rm part}$, and the spectator, $N_{\rm spect}$, respectively for nucleons which undergo at least one inelastic interaction and the non-interacted nucleons.
For a nucleus with mass number $A$, then $N_{\rm part} = A - N_{\rm spect}$,
which in the case of nucleus-$A$--nucleus-$B$ collision
depends on the impact parameter $b$ in the following form:
\begin{eqnarray}
N_{\rm part}^{AB}(b) &=& \int d^{2} s\, n_{\rm part}^{AB}({\bf b},{\bf s})\nonumber\\ &=&
A \int d^{2} s T_{A}({\bf s})
\left\{1-\left[1-\sigma_{in}  T_{B}({\bf b} - {\bf s})\right]^B \right\}\ \nonumber\\&&+
B \! \int d^{2} s T_{B}( {\bf b} - {\bf s})
\left\{1\!-\!\left[1\!-\!\sigma_{in} T_{A}({\bf s})\right]^{A} \right\}
\label{npartAB}
\end{eqnarray}
where $T_{A}({\bf s})= \int_{-\infty}^\infty d z
\rho_A(z,{\bf s})$ is the nuclear thickness function, normalized by  $\int d^2s\, T_A({\bf s}) = 1$.
The value of $\sigma_{in}$ represents the inelastic cross section
for the proton-proton interaction.

In the case of proton--nucleus ($pA$) collision, with the assumption of a point-like size of incident proton, one can put
$B = 1$ in \eqref{npartAB} and derive the expression for the number of participants and its average value in the form:
\begin{eqnarray}
N_{\rm part}^{pA}(b) &=&
A \sigma_{in} T_{A}( b) + \left\{1-P_{0}^{pA}(b) \right\},
\label{npartpA}\\
\langle N_{\rm part}^{pA} \rangle &=& \frac{\int b N_{\rm part}^{pA}(b)}
{\int b  [1-P_0(b)]} =
A \frac{\sigma_{in}}{\sigma_{pA}} + 1,
\label{meanNppA}
\end{eqnarray}
where $P_{0}^{pA}(b)$ is the probability that no any collision occurs between a proton and a nucleus at given impact parameter $b$, and the
Eq. \eqref{npartAB}.

One of the main characteristics of the interaction is the multiplicity of charged particles $N_{\rm ch}$, which can be related from the number of participants, $N_{\rm part}(b)$.
Indeed, one can obtain the form for the actual multiplicity, which
fluctuates around its mean value, $(2 \pi a \langle N_{\rm ch}(b)\rangle)^{-1/2}
e^{- \frac{[N_{\rm ch}-\langle N_{\rm ch}(b)\rangle]^2}
{2a\langle N_{\rm ch}(b)\rangle }}.
$
The coefficient $a$ fixes the width of such fluctuation.
%Hence, the differential cross section for nucleus-nucleus collision is given by following:
%\begin{equation}
%\frac{d\sigma_{mb}}{dN_{\rm ch}}=\int d^2b
%{\mathcal P}(N_{\rm ch}, N(b) )  \left[1-P_0(b) \right],
%\label{minbias}
%\end{equation}
%where $N(b)\equiv q N_{\rm part}(b)$ with constant parameter $q$, $P_0(b)=[1-\sigma_{in} T_{AB}(b)]^{AB}$
%is the probability that no any collision will occur, and $T_{AB}$ is the overlap parameter.
%Integration over all  charged particle multiplicity $dN_{\rm ch}$ gives us the total  cross section for nucleus-nucleus interaction:
%\begin{equation}
%\sigma_{AB} = \int dN_{\rm ch}  \frac{d\sigma_{mb}}{dN_{\rm ch}}=
%\int d^2b  \left[1-P_0(b) \right].
%\end{equation}

Let us introduce the unintegrated gluon distribution $\phi(x, k_t^2)$ which describes the
probability to find a gluon with a given $x$ and transverse momentum $k_t$
inside the nucleus $A$.
The main expression that can be used in order to obtain the inclusive production cross section from Ref. \cite{Kharzeev}:
\begin{equation}
\!E {d \sigma \over d^3 p} \!=\! {3\pi \over 2 p_t^2}\!
 \int^{p_{t}}\!\!d k_t^2
\alpha_{s}  \varphi_{A_1}(x_1, k_t^{2})\ \varphi_{A_2}(x_2, (p-k)_t^{2}),
\label{gencross}
\end{equation}
where, $\varphi_{A_{1},A_{2}} (x, k_t^{2})$, for $x_{1,2} = (p_{t}/\sqrt{s}) \exp(\mp y)$, for $y$ denoting the rapidity, is the gluon distribution of a nucleus, being $s$ the center-of-mass energy, involving two nuclei, $A_1$ and $A_2$.
Integrating Eq. \eqref{gencross} over $p_t$ yields the multiplicity distribution,
\begin{equation}
\label{MULTI}
\frac{d N}{d y} =  \frac{1}{\upsigma} \int  d^2 p_t E {d \sigma \over d^{3} p}
\end{equation}
with $\upsigma$ being the inelastic cross section.
The saturation scale, $Q_s$,  of deep inelastic scattering, reads
\begin{equation}
\label{QS}
Q^2_s(x)  =  Q^2_0 \left(\frac{x_0}{x} \right)^{\lambda},
\end{equation}
where the value of $\lambda = 0.288\pm0.03$.
Denoting $\mathcal{E}$ the collision energy,
then the energy and rapidity depend on the saturation scale as
\begin{equation}
\label{QSWY}
Q^2_s(A,y,\mathcal{E}) = Q_0^2(A,\mathcal{E}_0) \left( \frac{\mathcal{E}}{\mathcal{E}_0} \right)^{{\lambda}} e^{{\lambda} y} ,
\end{equation}
where
\begin{equation}
{\lambda} \equiv \frac{d\log(Q_s^2(x)/\Lambda^2_{\rm QCD})}{d\log (1/x)}\approx  0.252.\label{f1f}
\end{equation}
Integration of
\eqref{QSWY} yields \cite{Kharzeev}:
\begin{equation}
\label{QSR}
Q^2_s(\mathcal{E}) = \Lambda^2_{\rm QCD}e^{\sqrt{2 \delta \log\left(\frac{\mathcal{E}}{\mathcal{E}_0}\right) + \log^2\left[Q^2_s\left(\frac{\mathcal{E}_0}{\Lambda^2_{\rm QCD}}\right)\right]}}.
\end{equation}
In Eq. \eqref{QSR}, $Q^2_s(\mathcal{E}_0)$ denotes the saturation scale, characterized by the energy $\mathcal{E}_0$,
the parameter $\Lambda^2_{\rm QCD} = 0.04 {\rm GeV}^2$
and  $\delta = \lambda   \log(Q^2_{s0}/\Lambda^2_{\rm QCD})$ \cite{Kharzeev}.

At high energy range, the expression that links such concepts as
the energy, rapidity, and
atomic number dependence on hadron multiplicity reads \cite{Kharzeev}
\begin{eqnarray}
\varrho(s,y,\lambda)&=& N_{\rm part}\ \left({s \over s_0}\right)^{\lambda \over 2}\!\!\! e^{- \lambda |y|}
\left[\log\left({Q_s^2 \over \Lambda_{\rm QCD}^2}\right) - \lambda |y|\right]\nonumber\\
&&\times \left[ 1 +  \lambda |y| \left( 1 - {Q_s \over \sqrt{s}}\ e^{(1 + \lambda/2) |y|} \right)^4 \right].
\label{finres}
\end{eqnarray}

Eq. (\ref{finres}) reasonably predicts
the experimentally observed  hadron multiplicity at RHIC \cite{Kharzeev}, describing the energy dependence of the charged multiplicity in central $Au-Au$ collisions, at $\sqrt{s} = 130$ GeV and $\lambda = 0.25$
\cite{Kharzeev}. In the next section the CE will be computed, for the hadron multiplicity as the localized function\footnote{For fixed, but arbitrary, rapidities, as a function of the parameter $\lambda$.}. We will show that this precise value $\lambda = 0.253$, (\ref{f1f}), correspond to a global minimum of the CE, for central collisions, then corroborating to experimental values.

\section{Hadron multiplicity and configurational entropy}

First, remember that the CE concept involves localized functions \cite{Gleiser:2012tu,Gleiser:2013mga}. Therefore, we compute the Fourier transform of the energy--weighted correlation for the corresponding multiplicity distribution at the LHC, in the CGC approach. This can be implemented by using Eq. (\ref{finres}) as the localized function to be employed, for fixed rapidities:
\begin{equation}
\label{34}
{\varrho}({k,\lambda})=\frac{1}{2\pi}\! \int_{\mathbb{R}}\varrho (s,\lambda)\, e^{iks} d s.
\end{equation}  Therefore, the modal fraction reads
\begin{equation}\label{modall}
f_{\varrho({k,\lambda})}=\frac{\vert \varrho({k,\lambda}) \vert^2}{\int_{\mathbb{R}}\vert {\varrho({k,\lambda})\vert ^2} dk}.
\end{equation}
Using the corresponding formula for the CE \cite{Gleiser:2012tu}, one can get as following:
\begin{equation}
\label{333}
{\rm CE}(\lambda) =  - \int_{\mathbb{R}} f_{\varrho({k,\lambda})} \log  f_{\varrho({k,\lambda})} d k.
\end{equation}

\begin{figure}[!htb]
       \centering
                \includegraphics[width=2.9in]{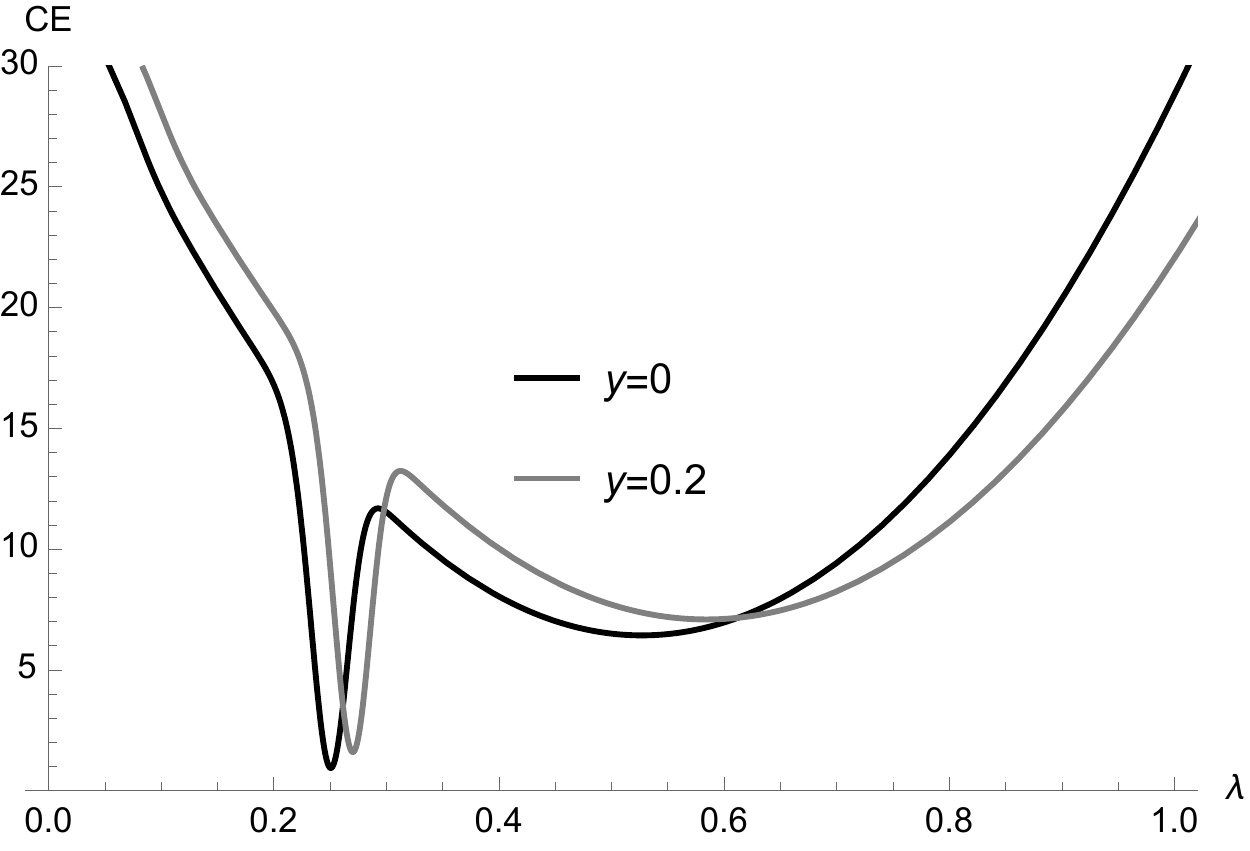}
                \caption{Configurational entropy (CE) as a function
                of the  saturation scale $\lambda$, for rapidity $y=0$ (black line) and $y=0.2$ (gray line). For $y=0$, the CE global minimum occurs at $\lambda=0.253$, whereas for $y=0.2$, the CE global minimum is at $\lambda=0.271$.}
                \label{fff1}
\end{figure}

The CE can be computed via Eqs. (\ref{34} - \ref{333}) for the
hadron multiplicity distribution at the LHC based on the CGC approach \cite{Kharzeev}, using Eq. (\ref{finres}). This is implemented numerically, into the plots in Fig. \ref{fff1}.

The results obtained for CE, for the multiplicity distribution, show an excellent agreement for the predicted  saturation scale $\lambda\approx 0.25$. It is worth to emphasize that  the
value $y=0$, adopted in Ref. \cite{Kharzeev} for the rapidity,  for the $Au$ nucleus at fixed energy of $\mathcal{E}_0$ = 130 GeV,  corresponds to the cut
of 0 - 6\% of most central collisions. Therefore, in the plots of Fig. \ref{fff1}, the only one to be compared to the literature will be $y=0$, being the another plot, regarding $y=0.2$, shown just for
the sake of completeness, as there is no related experimental data for $y=0.2$ in the literature, up to our knowledge. We will discuss more about it later.
%Q2
%s0 = 2 GeV2

Numerically calculated by Eqs. (\ref{34} - \ref{333}), using Eq. (\ref{333}), the nuclear CE  is then plot in Fig. \ref{fff1}.

The CE has a global minimum at $\lambda=0.253$ for $y=0$, whereas for $y=0.2$, the CE global minimum is at $\lambda=0.271$. These minima occur after a sharp decrement of the CE into a valley of more stable configurations. For $y=0$, this sharp valley of the CE is in the range $0.2\lesssim\lambda\lesssim0.299$. For $y=0.2$, the valley of the CE appears in the range $0.23\lesssim\lambda\lesssim0.32$.   
These results for $y=0$ match the expression (\ref{f1f}) and the ones in Ref.  \cite{Kharzeev}, involving  central $Au-Au$ collisions, at $\lambda = 0.252$,  within 0.39\%, for $y=0$, fixing  
the most stable configuration attained by the nuclear system.  
Using the concept of the Shannon's information entropy \cite{Gleiser:2011di}, one can figure out the critical points of the CE, that are global in the range analyzed, and thus establish the natural selection of the  saturation scale $\lambda$. It corresponds to the point where the nuclear system is more stable, from the informational point of view of the CE.
Besides, the only physically acceptable value adopted for the rapidity is $y=0$ \cite{Kharzeev}, for the $Au$ nucleus, since the value of the saturation scale, for $y=0$ was implemented in Ref. \cite{Kharzeev}. However, it is interesting to realize that the CE for $y=0$, at the minimum $\lambda=0.253$, is 0.92, whereas the  CE for $y=0.2$,  at the minimum $\lambda=0.271$, is 1.51. Since the value of the CE at the absolute minimum is lower for $y=0$, it means that the
set of modes constituting the nuclear system has a more stable configuration for $y=0$. We can show numerically that the higher the rapidity, the higher the CE is.

Therefore our results yield
a framework in nuclear physics that corroborates with the predicted values of the saturation scale $\lambda$, for any value of the rapidity, as a global minimum of the CE.
Hence, one can assert that the global minimum of the CE complies to the most dominant state of the nuclear configuration,  involving  central $Au-Au$ collisions.

\section{Conclusions}

Based on the CGC theory, the parton saturation results used in order to calculate the dependence of the hadron multiplicity on the rapidity, the energy and the saturation scale. Employing Eqs. (\ref{34} - \ref{333}),  the global minima of CE were computed for different values of rapidity. To compare with the literature, for the value $y=0$ the resulting saturation scale $\lambda=0.253$ matches phenomenological data within 0.39\% of precision. The minimum of the  CE predicts the predominant nuclear states, providing the natural set of the observables and show an excellent agreement not only with theoretical and phenomenological predictions but also with experimental data.

From the systematic analysis and detailed calculations, we conclude that in the framework of the CE,
the hadron multiplicity distribution dependence of the predicted saturation scale at $\lambda \approx 0.25$ provides an excellent description of the observed phenomenological data. Such calculations were obtained taking into account the fixed value of the rapidity $y=0$ for the central nucleus--nucleus collisions at the energy $\mathcal{E}_0$ = 130 GeV, which is shown in Fig. \ref{fff1} and which has been considered as the most appropriate data for the given value of rapidity in Ref. \cite{Kharzeev}.
The calculation at $y=0$ was compared with the appropriate system for $y=0.2$, showing that the value of the CE for the absolute minimum at $y=0$ reflects the more stable configuration of the nuclear system.

From the calculation, we found that the CE displays minima at $\lambda=0.253$ for $y=0$, and at $\lambda=0.271$ for $y=0.2$. The first result ($y=0$) is in agreement with the data observed in Ref. \cite{Kharzeev}, for the central nucleus--nucleus collisions and there is no experimental result yet for the second one ($y=0.2$), in the literature.
It should be noted that the critical points of the CE can be observed as the most predicted choice for the experimentally obtained value of the saturation scale $\lambda$. Thus, the minima on the calculation curve reflect the stability of the localized nuclear system.

One can study other types of nuclear configurations, with other field theoretical effects and other wavefunctions, as the ones proposed  in Refs. \cite{Bernardini:2012sc,daRocha:2005ti,Correa:2015vka,Correa:2016pgr,Bazeia:2013usa,daRocha:2011yr}. It is our aim to implement also the CE in such a context. 
  \acknowledgements
GK thanks to FAPESP (grant No.  2018/19943-6), for partial financial support. This paper is dedicated to the memory of CKR. 

\end{document}